\documentstyle[aps]{revtex}
\begin{document}
\draft

\def\ket#1{|#1\rangle}
\def\bra#1{\langle#1|}
\def\scal#1#2{\langle#1|#2\rangle}
\def\matr#1#2#3{\langle#1|#2|#3\rangle}
\def\keti#1{|#1)}
\def\brai#1{(#1|}
\def\scali#1#2{(#1|#2)}
\def\matri#1#2#3{(#1|#2|#3)}
\def\bino#1#2{\left(\begin{array}{c}#1\\#2\end{array}\right)}

\title{A microscopic framework for dynamical supersymmetry in nuclei}
\author{
Pavel Cejnar$^{1,2,}$\thanks
{Electronic address: pavel.cejnar@mff.cuni.cz}
and
Hendrik B. Geyer$^{1,}$\thanks
{Electronic address: hbg@sun.ac.za}
}
\address{
$^1$Institute of Theoretical Physics, University of Stellenbosch,
7602 Matieland, South Africa\\
$^2$Institute of Particle and Nuclear Physics, Charles University,
V Hole\v sovi\v ck\'ach 2, 180\,00 Prague, Czech Republic
}
\date{\today}
\maketitle

\begin{abstract}
We discuss and explore new aspects of the generalized Dyson mapping
of nuclear collective
superalgebras composed of an arbitrary fermion-pair algebra and
a set of single-fermion creation/annihilation operators. It is
shown that a direct consequence of the particular mapping procedure
is the conservation of the total number of ideal particles in the
resulting boson-fermion system. This provides a microscopic
framework for the phenomenological supersymmetric models based
on the U(6/$2\Omega$) dynamical superalgebras. Attention is paid to
the mapping of single-fermion creation and annihilation operators
whose detailed form cannot be determined on the phenomenological level.
We derive the general expansion of the single-fermion images that
result from the similarity transformation employed to ensure
non-redundant bosonisation in the ideal space. The method is then
illustrated in an application to the SO(4) collective algebra,
a natural extension of the SU(2) seniority model.
\end{abstract}

\section{Introduction}
\label{introduction}

It is well documented that low-energy collective states in even-even
nuclei can be successfully described using the interacting $s,d$-boson
models, the so-called IBM \cite{Iachello1}. The dynamical groups of
these models are bosonic unitary groups, either U$^{\rm B}$(6), or
its extension ${\rm U}^{{\rm B}\pi}(6)\otimes{\rm U}^{{\rm B}\nu}(6)$
if proton and neutron collective degrees of freedom are to be
treated separately. Similarly, the odd-$A$ and odd-odd nuclei
are described by the interacting boson-fermion models (IBFM)
\cite{Iachello2} with dynamical groups of the type
${\rm U}^{\rm B}(6)\otimes{\rm U}^{\rm F}(2\Omega)$
and ${\rm U}^{\rm B\pi}(6)\otimes{\rm U}^{\rm F\pi}(2\Omega_{\pi})
\otimes{\rm U}^{\rm B\nu}(6)\otimes{\rm U}^{\rm F\nu}(2\Omega_{\nu})$,
where $2\Omega$ is the capacity of the valence proton or neutron
shell and ${\rm U}^{\rm F}(2\Omega)$ the corresponding fermionic
unitary group. As a natural generalization of these approaches,
the above product boson-fermionic groups can be embedded into the
U(6/2$\Omega$) or ${\rm U}^{\pi}(6/2\Omega_{\pi})\otimes{\rm U}^{\nu}
(6/2\Omega_{\nu})$ supergroups \cite{Iachello3,Balantekin,Isacker}.
The immediate consequence of this step is the possibility of a
simultaneous
description of a given even-even nucleus with its odd-$A$ and odd-odd
neighbours---a consequence that was recently verified experimentally
\cite{Metz,Metz2,Groger} in the quartet of $^{194,195}$Pt and
$^{195,196}$Au nuclei.

A crucially important question imposed by the wide-ranging success of
all these phenomenological algebraic approaches \cite{Frank} concerns
their microscopic foundation. A variety of methods \cite{Klein} have
been developed with the aim to map the original fermionic
problem of an even number of particles into the bosonic language.
To understand the nature of the supersymmetric boson-fermion models,
however, the mapping must be extended to cover also the odd-fermion
degrees of freedom. In spite of numerous technical difficulties
it seems now that a basic understanding arises
\cite{Kaup,Dobaczewski1,Navratil1,Navratil2} of why the
U(6/2$\Omega$)-based dynamical supersymmetry is relevant in atomic
nuclei. A particularly promising approach, based on a superalgebraic
extension of the so-called Dyson mapping of fermion algebras
\cite{Klein}, was pioneered by Navr{\' a}til, Geyer, and Dobaczewski,
see Refs.~\cite{Dobaczewski1,Navratil1,Navratil2}. Among the main
advantages of the technique proposed there belongs the direct
relation of the resulting bosons to real fermionic pairs and the
conservation of two-body character of the model hamiltonian.
The method achieves its non-redundant bosonization as a two-step
process---no simpler construction is so far known---which requires
the utilization of a particular similarity transformation and
leads to typical Dyson-like non-hermitian structures.

The aim of the present paper is to review and extend the main
methodological aspects of the generalized Dyson mapping
\cite{Dobaczewski1,Navratil1,Navratil2} of nuclear collective
superalgebras, emphasising those features that are directly
related to phenomenological supersymmetric models.
In particular, we show that the conservation of the
total number of bosons plus fermions in phenomenological
models is a direct consequence of the mapping
procedure. We also discuss the general structure of
single-fermion transfer operators that cannot be deduced in detail on
the phenomenological level. Concrete new results are derived for
the mapping of the SO(4) collective algebra which had been used in
the first attempt \cite{Kaup} to investigate a possible microscopic
justification for the phenomenological supersymmetry in nuclei.

The plan of the paper is as follows: Notation and the general
collective superalgebra of fermion operators are introduced in
Sec.~\ref{superalgebra}. In Sec.~\ref{mapping} and \ref{simila}
we sketch the method of the fermion-boson mapping and subsequent
similarity transformations. The structure of the mapped
hamiltonian and properties of the single-fermion images are
then discussed in Sec.~\ref{conserva} and \ref{single}.
In Sec.~\ref{examples} we finally turn to some examples, based
on a simplified single-$j$ shell model, in particular to the
mapping of SU(2) and SO(4) collective superalgebras. We show how
matrix elements for the single-particle transfer (relevant to the
experimental identification of supersymmetry in nuclei
\cite{Metz,Metz2,Groger}) can be calculated in the ideal
boson-fermion space.

\section{Collective superalgebra}
\label{superalgebra}

Let us consider a fermionic system (atomic nucleus) with the Hilbert
space generated by a successive application of a finite number of
single-fermion creation operators $a^{\mu}$ to the physical vacuum
$\ket{0}$. We assume $\mu=1\dots 2\Omega$, enumerating the available
single-particle states (their number is even due to the Kramers
degeneracy) and denote the corresponding single-fermion
annihilation operators as $a_{\mu}\equiv(a^{\mu})^+$. Any physical
observable can then be obtained in terms of operators of the
following form,
\begin{equation}
A^{\mu_1\dots\mu_m}_{\nu_1\dots\nu_n}\equiv
a^{\mu_1}\dots a^{\mu_m}a_{\nu_1}\dots a_{\nu_n},
\label{gen}
\end{equation}
where $m,n=0,1,2\dots$ As fermionic operators satisfy the familiar
anticommutation relations,
\begin{equation}
\{a^{\mu},a_{\nu}\}=\delta^{\mu}_{\nu},\quad
\{a^{\mu},a^{\nu}\}=\{a_{\mu},a_{\nu}\}=0,
\label{fermion}
\end{equation}
the algebraic structure formed by the operators (\ref{gen}) is clearly
{\em not} an ordinary dynamical algebra \cite{Barut}. However, by
dividing the set of these operators to two subsets, {\em even\/} and
{\em odd}, according to whether the difference $m-n$ of the number
of creation and annihilation operators in
$A^{\mu_1\dots\mu_m}_{\nu_1\dots\nu_n}$ is even or odd, respectively,
we get the following schematic relations for commuting and anticommuting
operators from the two sectors:
\begin{equation}
[{\rm even},{\rm even}]={\rm even},\quad
[{\rm even},{\rm odd}]={\rm odd},\quad
\{{\rm odd},{\rm odd}\}={\rm even}.
\label{super}
\end{equation}
This means that the operators in Eq.~(\ref{gen}) define a
{\em superalgebra} \cite{Cornwell,Muller}. Indeed, the
mathematics of supergroups must be naturally involved in any
fermionic many-body problem if treated in the algebraic
framework \cite{Barut}.

The superalgebra of operators in Eq.~(\ref{gen}) can be reduced
to an ordinary dynamical algebra if one deals only with {\em even}
numbers ($N$) of particles. Any initial state is then represented by
an appropriate superposition of terms $A^{\mu_1\dots\mu_N}\ket{0}$
and transition operators may only contain terms with $m-n=0,2,4\dots$
(the many-body self hamiltonian is composed of terms with $m-n=0$).
Only the even sector of the set in Eq.~(\ref{gen}) is thus invoked.
Yet further crucial simplifications can be achieved if stronger
restrictions are imposed on the dynamics of the system. In nuclear
physics, the most important terms of the hamiltonian and transition
operators are often expressed via a certain limited set of fermion
pairs, briefly {\em bifermions}. We assume these pairs being
represented by the following most general creation operators,
\begin{equation}
A^i=\frac{1}{2}\,\chi^i_{\mu\nu}a^{\mu}a^{\nu}
\label{bifer}
\end{equation}
($i=1\dots M$; Greek indices occurring twice are subject to
summation over the whole range from 1 to $2\Omega$), where the
coefficients satisfy natural conditions $\chi^i_{\mu\nu}=
-\chi^i_{\nu\mu}=(\chi_i^{\mu\nu})^*$. [The bifermion annihilation
operators then read as $A_i\equiv(A^i)^+=\frac{1}{2}\chi_i^{\mu\nu}
a_{\nu}a_{\mu}$.] According to the concrete set of pairs
we choose, the operators $A^i$ and $A_i$ belong to a particular
{\em collective algebra\/} of the specific nuclear model. Note that
although we usually assume that the dynamics selects a limited
set of relevant pairs only, we can in principle consider all
possible pairs, $M=\Omega(2\Omega-1)$, with $i\mapsto(\mu,\nu)$,
$A^{\mu\nu}=a^{\mu}a^{\nu}$. The bifermion algebra is then identified
with SO(2$\Omega$) \cite{Dobaczewski1,Navratil1}.

The bifermion states  $A^i\ket{0}$ and $A^j\ket{0}$ are assumed
to be orthogonal and normalized to a common factor,
\begin{equation}
\frac{1}{2}\,\chi^{\mu\nu}_i\chi_{\mu\nu}^j=g\delta_i^j,
\label{norma}
\end{equation}
as typically follows from a diagonalization of the two-particle
problem.
This condition ensures the closure relations for the collective
algebra formed by the set of bifermion creation and annihilation
operators, $A^j$ and $A_i$, and by their commutators,
\begin{equation}
[A_i,A^j]=g\delta_i^j-\chi^j_{\sigma\mu}\chi_i^{\sigma\nu}
a^{\mu}a_{\nu},
\label{nebos}
\end{equation}
with $i,j=1\dots M$. (Note that $[A^i,A^j]=[A_i,A_j]=0$.) The
closure relations read as follows (the summation convention is
used for Latin indices),
\begin{eqnarray}
\left[\left[A_i,A^j\right],A_k\right] & = &
\left[\left[A^i,A_j\right],A^k\right]^+=c^{jl}_{ik}A_l,
\label{closure1}\\
\left[\left[A_i,A^j\right],\left[A_k,A^l\right]\right] & = &
c^{lj}_{km}\left[A_i,A^m\right]-c^{lm}_{ki}\left[A_m,A^j\right],
\label{closure2}
\end{eqnarray}
where the structure constants
\begin{equation}
c^{jl}_{ik}=\frac{1}{g}\
\chi_i^{\alpha\beta}\chi^j_{\beta\gamma}
\chi_k^{\gamma\delta}\chi^l_{\delta\alpha}
\label{struct_coll}
\end{equation}
satisfy symmetry relations
\begin{equation}
c^{jl}_{ik}=c^{jl}_{ki}=c^{lj}_{ik}=(c^{ik}_{jl})^*.
\label{symmetry}
\end{equation}
Note that according to Eq.~(\ref{closure2}) the commutators
(\ref{nebos}) form a core subalgebra of the collective algebra.

Of course, the use of the above collective algebra as the
approximate nuclear dynamical algebra can only be possible for
even nuclei. In the general case one has to consider also some
odd operators. In the following, we keep the collective algebra
of the above bifermion operators and extend it by considering
the single-fermion creation and annihilation operators that
give rise to single-fermion transfer operators between even
and odd systems. The algebra of collective operators forms the
even sector of the resulting superalgebra while the single-fermion
creation and annihilation operators belong to the odd sector.
Indeed, in agreement with the general superalgebraic rules
(\ref{super}), we have
\begin{eqnarray}
\left[A^i,a_{\mu}\right] & = & \left[a^{\mu},A_i\right]^+
=\chi^i_{\nu\mu}a^{\nu},
\label{sup1} \\
\left[A^i,a^{\mu}\right] & = & \left[a_{\mu},A_i\right]^+ =0,
\label{sup2} \\
\left[\left[A_i,A^j\right],a^{\mu}\right] & = & \left[a_{\mu},
\left[A_j,A^i\right]\right]^+ =\chi_i^{\mu\sigma}
\chi^j_{\sigma\nu}a^{\nu}.
\label{sup3}
\end{eqnarray}

Eqs.~(\ref{fermion}), (\ref{closure1}), (\ref{closure2}), and
(\ref{sup1})--(\ref{sup3}) define the superalgebra subject
to study in this paper. In fact, it is a combination of the
above collective algebra with the Heisenberg-Weyl superalgebra
\cite{Cornwell}. We will call it the {\em collective
superalgebra}.

\section{Fermion-boson mapping}
\label{mapping}

The superalgebraic nature of a general fermionic many-body problem
can be made explicit in terms of the usual bosonic and fermionic
degrees of freedom by using fermion-boson mapping techniques
\cite{Klein}. In this way, the actual (real) fermionic Hilbert
space is mapped onto an \lq\lq ideal\rq\rq\ space that describes
a system of a certain number of {\em bosons\/} and so-called
{\em ideal fermions}. To accomplish this task, a variety of
different approaches have been employed in the literature. In this
paper, we use a mapping technique that utilizes the
so-called Usui operator \cite{Usui,Donau,Dobaczewski1}, which is
closely related to the use of coherent and supercoherent states
\cite{Dobaczewski1}.

The Usui operator $T$ acts on the {\em product space\/} $\mathbf{H}=
\mathbf{H}_{\rm r}\otimes\mathbf{H}_{\rm i}$ of the real and ideal
Hilbert spaces. It transforms any real state vector $\ket{\psi}
\otimes\keti{0}$ (containing the ideal vacuum) into a corresponding
ideal state vector $\ket{0}\otimes\keti{\psi}$ (with the real
vacuum). If $P_0=\ket{0}\bra{0}\otimes 1$ and ${\cal P}_0=
1\otimes\keti{0}\brai{0}$ are projectors onto the real and
ideal vacua, respectively, one can define the {\em real\/} and
{\em ideal subspaces\/} of the product Hilbert space as ${\cal P}_0
\mathbf{H}$ and $P_0\mathbf{H}$ (they are isomorphic with the
original spaces $\mathbf{H}_{\rm r}$ and $\mathbf{H}_{\rm i}$).
(The rest of $\mathbf{H}$ is of no interest.) In the general case, the
Usui operator does not have to map the real subspace onto the
entire ideal subspace. The image of the real subspace, $T{\cal P}_0
\mathbf{H}$$\subset P_0\mathbf{H}$, forms the {\em physical subspace}
while the rest of the ideal subspace contains {\em spurious states}.

It seems reasonable to expect that any physically plausible mapping
should conserve
scalar products, i.e., must be unitary within the real and physical
subspaces. We will see, however, that this condition can be relaxed
without really loosing physical meaning of the mapping
\cite{Klein,Scholtz,Takada}. Let us consider the mapping of physical
operators, $O\mapsto\overline{O}$, defined through the requirement
$\overline{O}\,T=TO$, or equivalently
\begin{equation}
\overline{O}=TOT^{-1}
\label{oo}
\end{equation}
where $\overline{O}\equiv 1\otimes\overline{O}$ is the ideal image of
the real operator $O\equiv O\otimes 1$ and $T^{-1}$ is the inverse
Usui operator in the physical subspace. It is clear that any set of
operators within the real subspace is transformed into a set of
images acting in the ideal subspace, all the algebraic relations
(like $AB=C$, $A+B=C$, $[A,B]=C$, $\{A,B\}=C$\dots) being preserved
in the physical subspace (or in the overlap of definition ranges of
the operators involved with the physical subspace). If $T$ is
nonunitary, the mapping does not preserve properties related to the
hermitian conjugation. In particular, ideal images of general physical
operators will be nonhermitian. Nevertheless, because $T^{-1}$ must
exist within the physical subspace (this condition cannot
be relaxed!), all operator images remain isospectral with the
respective real operators and the eigenvectors are related by $T$.
Let us briefly recall that nonhermitian operators have two sets of
generally different eigenvectors, left and right: $\overline{O}
\keti{\psi^{\rm R}_i}=o_i\keti{\psi^{\rm R}_i}$ and
$\brai{\psi^{\rm L}_j}\overline{O}=\brai{\psi^{\rm L}_j}o_j$ and
different eigenspaces are not orthogonal but {\em biorthogonal}:
$\scali{\psi^{\rm L}_j}{\psi^{\rm R}_i}=0$ for $o_j\neq o_i$.
The nonunitarity of the mapping thus only induces the need
to treat separately right and left images of the physical
states according to the prescription
\begin{eqnarray}
\begin{array}{ll}
\keti{\psi^{\rm R}}=T\ket{\psi}\ ,&
\brai{\psi^{\rm L}}=\bra{\psi}\,T^{-1},\\
\brai{\psi^{\rm R}}=\bra{\psi}\,T^+,&
\keti{\psi^{\rm L}}=(T^{-1})^+\ket{\psi}\ .
\end{array}
\label{LR}
\end{eqnarray}
Note the hermitian conjugate of the Usui operator $T^+$ maps the
physical subspace back to the real space, but it is not identical
with $T^{-1}$ and, similarly, $(T^{-1})^+$ goes from the real
to physical space but does not coincide with $T$.

It should be stressed that to keep the mapping procedure
meaningful under these conditions, the proper distinction between
the physical and spurious subspaces is essential. In fact, any
operator that keeps the physical subspace invariant has an inverse
image in the real subspace while there may be no real counterpart
of operators acting within the entire ideal subspace.

Consider as the most trivial example a mapping that does nothing
but renames particles. We start with a set of real fermions
(created by $\{a^{\mu}\}_{\mu=1}^{2\Omega}$) and bosons
(created by $\{b^i\}_{i=1}^M$) and wish to end with a set of
ideal fermions ($\{\alpha^{\mu}\}_{\mu=1}^{2\Omega}$) and ideal
bosons ($\{B^i\}_{i=1}^M$). The Usui operator then reads
\begin{equation}
T=P_0\ \exp (B^ib_i+\alpha^{\mu}a_{\mu})\ {\cal P}_0.
\label{Utri}
\end{equation}
It is important to realize that the formal independence of physical
and ideal particles translates into the fact that any boson operator
commutes with all the other boson and fermion operators while the
real and ideal fermion operators {\em anticommute\/} with each other.
It is not difficult to see that under the operator in Eq.~(\ref{Utri})
any vector describing a state with fixed numbers of real particles
of the given types transforms into a vector with the same numbers
of the corresponding ideal particles. It means that the real subspace
is mapped onto the entire ideal subspace, keeping all scalar products
conserved and leaving no spurious states. The mapping (\ref{Utri})
is thus unitary within the real and ideal subspaces while vectors
orthogonal to the real subspace are annihilated by $T$. The operator
mapping corresponding to Eq.~(\ref{Utri}) is trivial: $b^i\mapsto B^i,
b_i\mapsto B_i, a^{\mu}\mapsto\alpha^{\mu}, a_{\mu}\mapsto\alpha_
{\mu}$. This enables one to construct the ideal image of any real
observable, conservation of the hermicity being guaranteed.

It is clear that the fermion-boson mapping we intend to perform
is not as trivial as the mapping in the previous example. Firstly,
the role of physical bosons is not be played by some actual
bosons but by fermion pairs from Eq.~(\ref{bifer}), whose
annihilation and creation operators do not really commute in the
bosonic way, see Eq.~(\ref{nebos}). Secondly, as bifermions are not
independent of single fermions, their operators do not commute with
fermion operators, see Eqs.~(\ref{sup1}) and (\ref{sup3}). In spite
of these difficulties, one still can keep the form of the Usui
operator from the previous example,
\begin{equation}
T=P_0\ \exp (B^iA_i+\alpha^{\mu}a_{\mu})\ {\cal P}_0,
\label{Urea}
\end{equation}
although some of its key properties differ from those discussed
above. In particular, the spurious sector of the ideal subspace
can no longer be avoided and $T$ ceases to be unitary even within
the real and physical subspaces. In fact, the actual justification
of Eq.~(\ref{Urea}) comes from the use of so-called supercoherent
states \cite{Fatyga} in both the real and ideal subspaces,
$\ket{C,\phi}\equiv\exp(C_iA^i+\phi_{\mu}a^{\mu})\ket{0}$ and
$\keti{C,\phi}\equiv\exp(C_iB^i+\phi_{\mu}\alpha^{\mu})\keti{0}$
(where $C_i$ and $\phi_{\mu}$ are complex and Grassman variables,
respectively). Any state $\ket{\psi}$ in the real space can
be represented by a function $f_{\psi}(C,\phi)\equiv
\scal{C,\phi\,}{\psi}$ and similarly any $\keti{\psi}$ in the
ideal space yields $g_{\psi}(C,\phi)\equiv\scali{C,\phi\,}
{\psi}$. It can be shown that the Usui operator from Eq.~(\ref{Urea})
conserves functional representations of the associated real and
ideal states \cite{Dobaczewski1}. [It should be stressed that not
every function $f(C,\phi)$ represents a real state $\ket{\psi}$
and those functional representations $g(C,\phi)$ in the ideal
space that have no counterpart in the real space constitute the
spurious sector. As the real and ideal supercoherent states span
the whole real and ideal spaces, respectively, $T$ does not map
the real and ideal supercoherent states to each other.]

Using the Baker-Campbell-Hausdorf formulas for
commuting the physical operators through the exponential in
Eq.~(\ref{Urea}), one can derive the following operator mapping
\cite{Dobaczewski1}:
\begin{eqnarray}
A^i\ \longmapsto\ & &{\cal A}^i+gB^i-\frac{1}{2}c^{ik}_{jl}B^jB^lB_k
-\chi^i_{\mu\sigma}\chi_j^{\nu\sigma}B^j\alpha^{\mu}\alpha_{\nu}
\nonumber\\
= & & {\cal A}^i+[{\cal A}_j,{\cal A}^i]B^j-\frac{1}{2}c^{ik}_{jl}
B^jB^lB_k\ ,
\label{opmap1}\\
A_i\ \longmapsto\ & &B_i\ ,
\label{opmap2}\\
a^{\mu}\ \longmapsto\ & &\alpha^{\mu}+\chi^{\mu\nu}_jB^j\alpha_{\nu}
\nonumber\\
= & & \alpha^{\mu}+[{\cal A}_j,\alpha^{\mu}]B^j\ ,
\label{opmap3}\\
a_{\mu}\ \longmapsto\ & & \alpha_{\mu}
\label{opmap4}
\end{eqnarray}
($i=1,\dots M$ and $\mu=1,\dots 2\Omega$). Here we introduced
ideal-bifermion operators ${\cal A}^i=\frac{1}{2}\,\chi^i_{\mu\nu}
\alpha^{\mu}\alpha^{\nu}$.
Note that---in agreement with the above discussion---the ideal
images of real creation and annihilation operators are not hermitian
conjugated. The mapping is nonunitary. In fact,
the ideal fermion and boson creation operators, $\alpha^{\mu}$
and $B^i$, have no inverse image in the real subspace (the
existence of these inverse images, $X$, would require the
fulfillment of the contradictory relations $\overline{X}{\cal P}_0
={\cal P}_0X$ with $\overline{X}=B^i$ or $\alpha^{\mu}$). Formulas
(\ref{opmap1}) and (\ref{opmap2}) can be compared to those derived
by mapping only the collective algebra without the odd sector
\cite{Dobaczewski3}. It turns our that the reduced Usui operator
$T=P_0\exp (B^iA_i){\cal P}_0$ leads to exactly the same images
of the bifermion operators $A^i$ and $A_i$ except that terms
associated with the ideal fermions are missing.

To complete the mapping of the whole fermionic superalgebra,
we need also images of the commutators of the bifermion
operators. These are given by the following formula,
\begin{eqnarray}
\left[A_i,A^j\right]\ \longmapsto\ & & g\delta^j_i-c^{jl}_{ik}B^kB_l-
\chi^j_{\mu\sigma}\chi_i^{\nu\sigma}\alpha^{\mu}\alpha_{\nu}
\nonumber\\
= & & [{\cal A}_i,{\cal A}^j]-c^{jl}_{ik}B^kB_l\ ,
\label{comap}
\end{eqnarray}
which can be obtained either by mapping the r.h.s. of
Eq.~(\ref{nebos}), or---in a simpler way---by commuting the images
of bifermion operators in Eqs.~(\ref{opmap1}) and (\ref{opmap2}).
Also the anticommutation relations
of real and mapped single-fermion operators are identical. This
means that Eqs.~(\ref{opmap1})--(\ref{comap}) indeed define an
equivalent ideal boson-fermion realization of the real fermion
superalgebra from Sec.~\ref{superalgebra}.

\section{Similarity transformations}
\label{simila}

\subsection{Hermitisation}
\label{similaa}

It was stressed above that the mapping in
Eqs.~(\ref{opmap1})--(\ref{comap}) is not unitary so that the ideal
images of real observables are generally nonhermitian. On the other
hand, we know that within the physical subspace the spectra of these
images are real-valued, identical with the spectra of physical operators.
For any particular physical ideal-image operator $\overline{O}$
it should therefore be possible to find a similarity transformation
$\overline{O}\mapsto\overline{O}'=S_{\rm H}\overline{O}S_{\rm H}^{-1}$
such that $\overline{O}'$ is hermitian. If $\keti{\psi^{\rm R}_i}$
and $\brai{\psi^{\rm L}_j}$ are sets of right and left eigenvectors
of $\overline{O}$, the operator $S_{\rm H}$ must satisfy
$\matri{\psi^{\rm L}_j}{S_{\rm H}^+S_{\rm H}}{\psi^{\rm R}_i}=
\delta_{ji}$, or, equivalently, $S_{\rm H}^+S_{\rm H}\overline{O}=
\overline{O}^+S_{\rm H}^+S_{\rm H}$. Indeed, one can take, for
example,
$S_{\rm H}=(TT^+)^{-1/2}$, where $T^+={\cal P}_0\exp(A^iB_i+a^{\mu}
\alpha_{\mu})P_0$ [cf. Eq.~(\ref{Urea})]. However, as shown by Kim
and Vincent
\cite{Kim}, it is often favorable to exploit the ambiguity of the
hermitisation transformation---it is determined up to an
arbitrary unitary transformation---to set constraints upon the
image of one of the observables, e.g., the hamiltonian $\overline{H}$.
Namely, if $S_{\rm H0}$ hermitizes the hamiltonian, then
\begin{equation}
S_{\rm H}=(S_{\rm H0}TT^+S_{\rm H0}^+)^{-\frac{1}{2}}S_{\rm H0}
\label{herm}
\end{equation}
hermitizes, within the physical subspace, all physical
observables, while retaining the prescribed form of the hamiltonian,
$S_{\rm H}\overline{H}S_{\rm H}^{-1}=S_{\rm H0}\overline{H}S_{\rm H0}
^{-1}$. This is very important since we naturally require
that the hermitisation does not spoil some important features
of the mapped hamiltonian, for instance, its one- plus two-body
character.

Hermitisation transformations preserving the two-body character
of the hamiltonian were indeed described in some particular
cases \cite{Kim}, but no general algorithm is known. One direct
approach is simply to guess the desired hermitian
operator $\overline{H}'$ isospectral with $\overline{H}$ and
to construct a consistent similarity transformation.
This is possible, under some specific conditions, using the
following expression,
\begin{equation}
S_{\rm H0}^{-1}=\sum_{k=0}^{\infty}\left(\frac{1}
{{\hat C}-C}P\right)^k_{\wedge}\ ,
\label{posi}
\end{equation}
where $P=\overline{H}-\overline{H}'$ and $C$ is any operator
satisfying $[C,P]=[\overline{H}',P]$. In Eq.~(\ref{posi}) we
introduce the notation in which the mark \lq\lq$_{\wedge}$\rq\rq\
indicates the position where the operator with hat (the first
$C$) is to be evaluated. The derivation of this formula and
the positional operator formalism are sketched in Appendix A.
It is important to stress that Eq.~(\ref{posi}) holds true
only for a non-degenerate spectrum of $\overline{H}$, while otherwise
divergence problems can be encountered.

In the majority of cases it is difficult (if not impossible)
to derive explicit expressions for the hermitized images of physical
operators using the general transformation in Eq.~(\ref{herm}).
At first this difficulty seems to put serious restrictions on
the use of the mapping technique described above. Fortunately,
the calculation of {\em matrix elements\/} of physical operators
can be performed without really knowing the hermitized images in
the operator form, by  using the obvious identity $\matr{\psi_1}{O}
{\psi_2}=\matri{\psi_1^{\rm L}}{\overline O}{\psi_2^{\rm R}}$
or its modification
\begin{equation}
\matr{\psi_1}{O}{\psi_2}=\sqrt{
\matri{\psi^{\rm L}_1}{\overline{O}}{\psi^{\rm R}_2}
\matri{\psi^{\rm L}_2}{\overline{O^+}}{\psi^{\rm R}_1}^*
},
\label{matrixele}
\end{equation}
that both directly result from Eqs.~(\ref{oo}) and (\ref{LR})
(the second identity is usually favoured in practical calculations
as we will see in Sec.~\ref{examplesd}). The
evaluation of the hermitisation transformation is turned here
into another nontrivial task---finding the left and right
images of general physical states. However, this can already be
accomplished for certain sets of states, namely those
generated by some creation operators from the real vacuum,
i.e., for states having the form $\ket{\psi}=X^+\ket{0}$
(where $X^+$ represents, e.g., a sequence of single-fermion
and/or bifermion creation operators). Then one can write
$\keti{\psi^{\rm R}}=\overline{X^+}\keti{0}$ and
$\brai{\psi^{\rm L}}=\brai{0}\,\overline{X}$ with
$\keti{0}\equiv\keti{0^{\rm R}}=\keti{0^{\rm L}}$.
In this way, one can evaluate---using only the
ideal images of state vectors and operators---the complete set
of matrix elements of the given real operator in an appropriate
real basis (the single-particle basis, for instance). The goal
of the mapping can thus be achieved \cite{Hahne,Takada2,Takada}.

\subsection{Bosonisation}
\label{similab}

The necessity for a similarity transformation following the
mapping described in the Sec.~\ref{mapping} appears even before
considering the hermitisation problem. This is evident from
Eq.~(\ref{opmap1}) where the ideal image of the real pair
creation operator contains the ideal pair creation operator,
$A^i\mapsto{\cal A}^i+R^i$ with $R^i=gB^i-\frac{1}{2}c^{ik}_{jl}
B^jB^lB_k-\chi^i_{\mu\sigma}\chi_j^{\nu\sigma}B^j\alpha^{\mu}
\alpha_{\nu}$. While all terms in $R^i$ translate the creation
of a bifermion in the real space into the creation of a boson
in the ideal space (this can be accompanied by an interaction
with another boson or fermion), ${\cal A}^i$ just introduces
an equivalent ideal fermion pair. The real pairs are thus
not truly bosonized by the mapping. In particular, the real
bifermion state $A^i\ket{0}$ is transformed into a superposition
of ideal bifermion and boson states, $({\cal A}^i+gB^i)\keti{0}$,
and real fermion-fermion interactions are exactly transmitted
to the ideal hamiltonian, where the additional boson and
boson-fermion terms (see Sec.~\ref{conserva}) only obscure
the original problem.

This difficulty can be again overcome with the aid of the
formalism sketched in Appendix A. Indeed, when
considering the operator $A^jA_j$ and its image
${\cal A}^jB_j+R^jB_j$, we see that the unwanted part
containing ${\cal A}^j$ does not affect the spectrum of
the image. This
follows from the fact that while the $R^jB_j$ operator is
diagonal in the basis characterized by numbers of ideal bosons
and fermions, the ${\cal A}^jB_j$ term has an upper off-diagonal
block structure in the same basis. We therefore anticipate the
existence of a similarity transformation $S_{\rm B}$ with the
following properties:
\begin{eqnarray}
S_{\rm B}({\cal A}^i+R^i)S_{\rm B}^{-1}=R^i\ ,
\label{simi0}\\
S_{\rm B}B_iS_{\rm B}^{-1}=B_i\ .
\label{simi1}
\end{eqnarray}
The form of $S_{\rm B}^{-1}$ is given by Eq.~(\ref{posiA}) in the
Appendix with $O'=R^jB_j$ and $P={\cal A}^jB_j$. However, it can
be shown \cite{Navratil1} that $[O',P]=[-{\cal C}_{\rm F},P]$,
where ${\cal C}_{\rm F}={\cal A}^l{\cal A}_l$ is the Casimir
operator of the ideal fermion core algebra, $[{\cal C}_{\rm F},
[{\cal A}_i,{\cal A}^j]]=0$, conserving the total number of
ideal fermions, ${\cal N}=\alpha^{\mu}\alpha_{\mu}$. In agreement
with Eqs.~(\ref{posi}) and (\ref{sumpos3}) we thus arrive at
\begin{equation}
S_{\rm B}^{-1}=\sum_{k=0}^{k_{\rm max}}\left(\frac{1}{{\cal C}
_{\rm F}-{\hat{\cal C}}_{\rm F}}{\cal A}^jB_j\right)^k
_{\wedge}=\exp{\left[\frac{{\cal N}-{\hat{\cal N}}}
{2({\cal C}_{\rm F}-{\hat{\cal C}}_{\rm F})}{\cal A}^jB_j\right]}
_{\wedge}\ ,
\label{simi2}
\end{equation}
where the upper bound of the sum, $k_{\rm max}$, reflects the
finiteness of the fermionic space. It is clear that $k_{\rm max}
\leq\Omega$ and that the real cut off depends on the numbers
${\cal N}$ and $N_{\rm B}$ of the ideal fermions and bosons present
in the state to be transformed (in this way also the higher-order
terms in expansion of the exponential naturally vanish). Let us
stress again that due to the limitations mentioned above and in
Appendix A, there is no general guarantee that Eq.~(\ref{simi2})
converges. This is further illustrated in Sec.~\ref{examples}, where the
convergence requirement will set some limits upon the states
to be transformed.

In order to obtain the transformed images of general physical
operators, we also need to determine the form of the inverse
similarity transformation $S_{\rm B}$. As the expansion of
$S_{\rm B}^{-1}$ in Eq.~(\ref{simi2}) consists of terms
which increase the number of ideal fermions by $\Delta{\cal N}
=2k=0,+2,+4,+6\dots$, the same must hold true also for the
$S_{\rm B}$. If we define
\begin{equation}
S_k=\left(\frac{1}{{\cal C}_{\rm F}-{\hat{\cal C}}_{\rm F}}
{\cal A}^jB_j\right)^k_{\wedge}\ ,
\label{s1k}
\end{equation}
i.e., if we rewrite Eq.~(\ref{simi2}) as
\begin{equation}
S_{\rm B}^{-1}=1+S_1+S_2+S_3+\dots\ ,
\label{sser1}
\end{equation}
we find that
\begin{equation}
S_{\rm B}=1+{\tilde S}_1+{\tilde S}_2+{\tilde S}_3+\dots
\label{sser2}
\end{equation}
with
\begin{equation}
{\tilde S}_k=\sum_{n=1}^k (-)^n\!\!\!\!\!\!
\sum_{k_1+k_2+\dots +k_n=k}\!\!\!\!
S_{k_1}S_{k_2}\dots S_{k_n}\ .
\label{s2k}
\end{equation}
In particular, ${\tilde S}_1=-S_1$, ${\tilde S}_2=-S_2+
S_1^2$, \dots, cf.~Ref.\cite{Navratil1}. These expressions enable
one to evaluate the similarity transformation
$S_{\rm B}XS_{\rm B}^{-1}$ of an arbitrary operator $X$, which
changes the number of ideal fermions by a specific value
$\Delta{\cal N}$, as a series where individual terms correspond to
$\Delta{\cal N}$, $\Delta{\cal N}+2$, $\Delta{\cal N}+4$, etc.
We use these expansions in Sec.~\ref{single} when discussing
the general form of transformed single-fermion images.

\section{Conservation of the number of ideal particles}
\label{conserva}

One of the most interesting questions immediately arising from the
previous considerations concerns the link to superalgebras of
the type U($M/2\Omega$) known from phenomenological boson-fermion models
of nuclear structure \cite{Iachello3,Balantekin,Isacker,Frank}. The
use of these dynamical superalgebras on the phenomenological level
is motivated by the fact that they provide a direct generalization
of the unitary bosonic and fermionic algebras that proved to be
relevant and successful in the description of collective states in both
even
and odd (odd-$A$ or odd-odd) nuclei \cite{Iachello1,Iachello2,Frank}.
In fact, generators of the proton-neutron superalgebra ${\rm U}^{\pi}
(6/2\Omega_{\pi})\otimes{\rm U}^{\nu}(6/2\Omega_{\nu})$ produce
a class of related hamiltonians that seems general enough to
simultaneously
describe low-energy spectra in quartets of nuclei whose nucleon
(proton and/or neutron) numbers differ by one \cite{Isacker,Frank}.

The key feature of the U($M/2\Omega$) superalgebras is that their
generators conserve the total number of bosons plus fermions,
$N_{\rm BF}=N_{\rm B}+{\cal N}$ (where $N_{\rm B}=B^iB_i$).
We thus enquire whether this also holds for the hamiltonian mapped
from a microscopic real-fermion hamiltonian.
Let us stress that this property cannot be deduced from the
conservation of the number of real fermions, $N$, by the original
nuclear hamiltonian since $N$ corresponds to
$2N_{\rm B}+{\cal N}$ on the boson-fermion level, as dictated by
the fermion-boson mapping [Eq.~(\ref{iman}) below]. It is
nevertheless clear from
Eqs.~(\ref{opmap1})--(\ref{comap}) that any fermionic many-body
hamiltonian composed of operators belonging to the collective
algebra, e.g.,
\begin{eqnarray}
H=u & + & v_i^j A^iA_j+w^i_j [A_i,A^j]=(u+gw^i_i)
\nonumber\\
& - & (\chi^j_{\mu\sigma}\chi_i^{\nu\sigma}w^i_j)a^{\mu}a_{\nu}+
\left(\frac{1}{4}\chi^i_{\mu\nu}\chi_j^{\pi\sigma}v_i^j\right)
a^{\mu}a^{\nu}a_{\sigma}a_{\pi}
\label{Hamr}
\end{eqnarray}
[where $v_i^j=(v^{i}_j)^*$ and $w_i^j=(w^{i}_j)^*$ are arbitrary
coefficients associated with two- and one-body interactions,
respectively, and $u=u^*$ is an additive constant], is mapped onto
an ideal hamiltonian that indeed conserves the total number of
ideal particles. This conclusion remains unchanged even after the
similarity transformation in Eqs.~(\ref{simi0}) and (\ref{simi1}).
The resulting hamiltonian keeps the same ideal-fermion mean field
as the original real hamiltonian (\ref{Hamr}), but the fermion-fermion
interaction is replaced by boson-involving terms that describe a
boson mean field and boson-boson plus boson-fermion interactions:
\begin{eqnarray}
H\ \longmapsto\ (u+gw^i_i) & - & (\chi^j_{\mu\sigma}
\chi_i^{\nu\sigma}w^i_j)\alpha^{\mu}\alpha_{\nu}
+(gv^j_i-c^{jk}_{il}w^l_k)B^iB_j
\nonumber\\
& - & \left(\frac{1}{2}c^{mk}_{ij}v^l_m\right)B^iB^jB_kB_l
-(\chi^k_{\mu\sigma}\chi_i^{\nu\sigma}v^j_k)
B^i\alpha^{\mu}B_j\alpha_{\nu}
\label{Hami}
\end{eqnarray}
The image in Eq.~({\ref{Hami}) is still non-hermitian in both
interaction terms, but, as discussed in Sec.~\ref{similaa}, the
similarity transformation $S_{\rm H}$ can be chosen such
that it does not affect the particle number conservation. (In
the matrix representation connected with fixed particle numbers
the hamiltonian has a block-diagonal structure which can be
preserved by a suitably selected transformation $S_{\rm H0}$.)
This result holds true for {\em any\/} collective algebra we
decide to start with.

The image (\ref{Hami}) conserves numbers of bosons and
fermions {\em separately}, which is the structure known
from the interacting boson-fermion model \cite{Iachello2}. It
seems therefore that ${\rm U}^{\rm B}(M)\otimes {\rm U}^{\rm F}
(2\Omega)$ could be equally well chosen as the dynamical algebra
on the phenomenological level instead of U($M/2\Omega$). Note
that the choice of the phenomenological dynamical algebra
(superalgebra) is more or less a matter of convenience; it
certainly does not result from the mapping procedure which only
constructs a boson-fermion realization of the original collective
superalgebra. If, nevertheless, the U($M/2\Omega$) dynamical
superalgebra is employed, it must be decomposed into the above
product of boson and fermion unitary algebras in the very first
step of any relevant dynamical symmetry chain. This indeed happens
in the phenomenological model \cite{Isacker} used to analyse
experimental data \cite{Metz,Metz2,Groger}. From this point of view,
the hitherto discussed supersymmetric description of neighboring
even-even, odd-odd, and odd-$A$ nuclei relies just on the use of
the IBFM with a single set of parameters, which is a
natural expectation based on the mapping of the same
microscopic hamiltonian (\ref{Hamr}) acting on spaces with
various real-fermion numbers (see also Ref. \cite{Navratil2} in this
regard).
At the same time we note that the above
considerations are not in contradiction with the recently proposed
possibility \cite{Jolos,Jolos2,Jolos3} that
a U($n/m$)$\subset$U($M/2\Omega$) supergroup may in fact constitute
a real invariance symmetry of the nuclear hamiltonian, without reference
to an underlying dynamical symmetry, giving rise to boson-fermion
\lq\lq supermultiplets\rq\rq\ in neighboring nuclei.

We conclude this Section by the remark that {\em general\/}
one- plus two-body hamiltonians (for instance those containing
general single-particle terms $\varepsilon^{\nu}_{\mu}a^{\mu}
a_{\nu}$) do {\em not\/} have to conserve the number of ideal
particles after the mapping. This probably misled the authors
of Ref.\cite{Kaup} who ascribed the conservation property to
only the Schwinger type of mapping while it was alleged to fail
for mappings which associate bosons with fermion pairs. However,
from the above discussion we see that the ideal-particle number
{\em is\/} indeed conserved in the generalized Dyson mapping as
far as the mapped collective algebra represents the dynamical
algebra of the fermionic hamiltonian [for instance, if the
single-particle terms are given only by the commutators in
Eq.~(\ref{nebos})]. An enquiry about the most general set of
fermion hamiltonians (beyond the preselected dynamical algebra)
that conserve $N_{\rm BF}$ after the mapping is hampered by the
following difficulties: (i) the ideal image of the hamiltonian
does not have to commute with $N_{\rm BF}$ in the whole ideal
space, but only in the physical subspace, and (ii) there are no obvious
candidates for fermion space counterparts to the observables
associated with $N_{\rm B}$ and ${\cal N}$.

\section{Single-fermion images}
\label{single}

While the action of the similarity transformation (\ref{simi2})
on the bifermion images is by construction guaranteed to yield the compact
results (\ref{simi0}) and (\ref{simi1}),
the expressions for transformed single-fermion
images can only be determined in an expanded form, using
Eqs.~(\ref{s1k})--(\ref{s2k}). Denoting the \lq\lq bare\rq\rq\
single-fermion images appearing in Eqs.~(\ref{opmap3}) or (\ref{opmap4})
by $X$, one can write
\begin{equation}
S_{\rm B}XS_{\rm B}^{-1}=X+X_1+X_2+X_3+\dots\ ,
\label{xxx}
\end{equation}
where individual terms are determined by
\begin{equation}
X_k=[X,S_k]+\sum_{n=2}^k(-)^{n-1}\!\!\!\!\!\!\!\!\!\!\!\!\!\!
\sum_{k_1+k_2+\dots+k_n=k}\!\!\!\!\!\!\!\!\!
S_{k_1}S_{k_2}\dots S_{k_{n-1}}[X,S_{k_n}]\ ,
\label{xk}
\end{equation}
or recursively from
\begin{eqnarray}
X_1 & = & [X,S_1]\ ,\qquad X_2=[X,S_2]-S_1X_1
\ ,\qquad\dots\ ,\nonumber\\
X_k & = & [X,S_k]-S_1X_{k-1}-S_2X_{k-2}-\dots -S_{k-1}X_1\ .
\label{recur}
\end{eqnarray}
In the expressions above $X$ can, in principle,
be any physical operator. With $X=\alpha_{\mu}$ the term
$X_k$ changes the number of ideal fermions by $\Delta{\cal N}=2k-1$,
while with $X=\alpha^{\mu}+\chi^{\mu\nu}_jB^j\alpha_{\nu}\equiv
X'+X''$ we have $\Delta{\cal N}=2k+1$ for $X'_k$ and
$\Delta{\cal N}=2k-1$ for $X''_k$. The transformed images of both
annihilation and creation operators thus contain terms with
$\Delta{\cal N}=-1,+1,+3,+5\dots$ \cite{Navratil1}.

The series (\ref{xxx}) for transformed single-fermion images
comprises (i) the operators contained in the bare
single-fermion images, i.e., $\alpha_{\mu}$ or $\alpha^{\mu}$ and
$\chi^{\mu\nu}_iB^i\alpha_{\nu}$, and (ii) those in the similarity
transformations $S_{\rm B}^{-1}$ and $S_{\rm B}$, i.e.,
${\cal A}^iB_i$ and $({\cal C}_{\rm F}-{\hat{\cal C}}_{\rm F})$.
In any term of the series, there can be only one operator from (i)
and an arbitrary combination (no restrictions to multiplicity)
of operators from (ii). To determine the physical interpretation
of these expressions by inspection, we first commute $({\cal C}_{\rm F}-
{\hat{\cal C}}_{\rm F})$ from all places of its occurrence to
the respective positional marks and then to the right-hand side
in all terms of the series. It then turns out that the resulting
formulas can be decomposed into building blocks representing
some elementary processes: (a) Processes corresponding to the
ideal-fermion creation,
\begin{eqnarray}
& & \left[{\cal C}_{\rm F},\left[{\cal C}_{\rm F}\dots
[{\cal C}_{\rm F},\alpha^{\mu}\right]..\right]_n
\ ,\nonumber\\
& & \left[{\cal C}_{\rm F},\left[{\cal C}_{\rm F}\dots
[{\cal C}_{\rm F},\chi^{\mu\nu}_iB^i\alpha_{\nu}\right]..\right]_n
\quad n=0,1,2\dots\ .
\label{procre}
\end{eqnarray}
Here, the $n=0$ terms emerge as just  $\alpha^{\mu}$ and $\chi^{\mu
\nu}_iB^i\alpha_{\nu}$, the $n=1$ terms as $\chi^{\mu\nu}_i{\cal A}
^i\alpha_{\nu}$ and $\chi^{\mu\nu}_i\chi^j_{\sigma\nu}\alpha
^{\sigma}{\cal A}_jB^i$, etc. These expressions can be interpreted
as processes that encompass single fermion creation, coupling of
the created fermion into a pair, and the bifermion--boson
transformations (all fermions of course being of the ideal type).
(b) Processes corresponding the ideal-fermion
annihilation,
\begin{equation}
\left[{\cal C}_{\rm F},\left[{\cal C}_{\rm F}\dots
[{\cal C}_{\rm F},\alpha_{\mu}\right]..\right]_n
\quad n=0,1,2\dots\ ,
\label{proani}
\end{equation}
which appear just as the hermitian conjugate of the first
commutator in Eq.~(\ref{procre}) and receive analogous
diagrammatic interpretations. (c) Background processes
accompanying (a) and (b),
\begin{equation}
\left[{\cal C}_{\rm F},\left[{\cal C}_{\rm F}\dots
[{\cal C}_{\rm F},{\cal A}^iB_i\right]..\right]_n
\quad n=0,1,2\dots\ ,
\label{background}
\end{equation}
that denote various forms of the boson decomposition into
ideal fermions and bifermions; for $n=0$ we have just
${\cal A}^iB_i$, while the $n=1$ term is $\chi^i_{\nu\pi}\chi_j
^{\nu\sigma}{\cal A}^j\alpha^{\pi}\alpha_{\sigma}B_i$, etc.
While expressions from Eq.~(\ref{procre}) or (\ref{proani})
Appear only once in each term of the series, those from
Eq.~(\ref{background}) generally have a multiple occurrence.

We have already seen that the most general transformed images of
the single-fermion creation and annihilation operators contain
terms that change the number of ideal fermions by $\Delta{\cal N}
=-1,+1,+3,+5\dots$. In view of the elementary processes (a)--(c),
the actual value of $\Delta{\cal N}$ in a given term is
determined by the number of repetitions of the background
processes (\ref{background}). Relative weights appearing
with increasing $\Delta{\cal N}$ are expected to
decrease according to the increasing power of the denominator
in Eq.~(\ref{s1k}) [cf. Eqs.~(\ref{sksu2}) and (\ref{skso4})
below]. In addition, terms corresponding to large
$\Delta{\cal N}$ are not likely to play a significant role
in matrix elements for low-energy nuclear states as the
decomposition of bosons into separate non-collective
fermions is associated with higher energy
excitations. As argued in Ref.\cite{Navratil1}, it may be
plausible to cut off the terms with $\Delta{\cal N}\geq+3$.
We will see in Sec.~\ref{examples} that for some algebras
these terms can vanish identically. Even with the restriction
$\Delta{\cal N}\leq+1$, however, the most general formula built
of terms from Eqs.~(\ref{procre})--(\ref{background}) comprises
an infinite series (terms with all $n$'s). The situation is much
simplified if ${\cal C}_{\rm F}$ is just a function of
${\cal N}$ (or number operators associated with some fermionic
subspaces). Then, evaluating only the $\Delta{\cal N}=\pm 1$
terms, one gets
\begin{eqnarray}
S_{\rm B}\alpha_{\mu}S_{\rm B}^{-1} & = &
\alpha_{\mu}+\chi^i_{\mu\nu}\alpha^{\nu}B_i+
{\cal A}^i\alpha_{\mu}B_iF({\cal N})+\dots
\label{ani}\\
S_{\rm B}(\alpha^{\mu}+\chi^{\mu\nu}_iB^i\alpha_{\nu})
S_{\rm B}^{-1} & = &
\chi^{\mu\nu}_iB^i\alpha_{\nu}+\alpha^{\mu}+
\chi^{\mu\nu}_i\chi^j_{\nu\sigma}\alpha^{\sigma}B^iB_j
\nonumber\\
& & \quad +\chi^{\mu\nu}_i{\cal A}^i\alpha_{\nu}F'({\cal N})+
\chi^{\mu\nu}_i{\cal A}^j\alpha_{\nu}B^iB_jF({\cal N})+\dots
\label{cre}
\end{eqnarray}
where $F({\cal N})$ and $F'({\cal N})$ are some functions of the
ideal-fermion number(s) that are directly related to the form
${\cal C}_{\rm F}=f({\cal N})$. This general result is
illustrated by specific examples in Sec.~\ref{examplesc}.

It is clear that even after the transformation (\ref{xxx}) the
images of real-fermion creation and annihilation operators are
not hermitian conjugated; cf.\ Eqs.~(\ref{ani}) and (\ref{cre}).
As the general hermitisation transformation, Eq.~(\ref{herm}),
of the single-fermion images in an operator form seems
intractable, at least in the general case, one has to turn
to the evaluation of the single-fermion matrix elements in
a specific basis by the method described in Sec.~\ref{similaa},
see Eq.~(\ref{matrixele}). This step, of course, critically
depends on the concrete form of single-fermion images after
the bosonisation transformation. Examples are given in
Sec.~\ref{examplesd}.

We conclude this Section by noting that the general results
discussed here may suggest expressions suitable for the determination
of single-nucleon transfer amplitudes within the phenomenological
superalgebraic models. In fact, on the phenomenological level only very
general considerations, such as tensorial properties and effective
particle
number properties, relate to this question, as the transfer operators
are beyond
the U$(M/2\Omega)$ superalgebra. A link to microscopic models
is thus essential. For instance, we can recall the formula fitted to
experimental data in Refs.\cite{Metz,Metz2} and compare it with the most
general form discussed above. We see immediately that although the
image of the single-fermion creation operator given by Eq.~(3) in
Ref.\cite{Metz} contains terms describing plausible processes with
$\Delta{\cal N}=+1$, some other possibly relevant terms are missing.

\section{Examples: SU(2) and SO(4) mappings}
\label{examples}

\subsection{Definition of the algebras}
\label{examplesa}

In this section, we illustrate the general considerations of the previous
sections
by simple examples concerning fermions in a single shell with
total angular momentum $j$ (half integer). Accordingly we
consider a set of $2\Omega=2j+1$ single-particle states created
by $a^{\mu}\equiv a^{\dagger}_{j\mu}$ with $\mu=-j\dots+j$. We
will deal with the simplest algebras based on these operators,
namely the SU(2) seniority algebra \cite{Greiner} and
the extended SO(4) algebra \cite{Matsuyanagi,Kaup}

In the SU(2) case we introduce only one type of fermion pair,
namely
\begin{equation}
A^1\equiv S^{\dagger}=\frac{1}{2}\sum_{\mu}(-)^{j-\mu}
a^{\mu}a^{-\mu}\ .
\label{Sdef}
\end{equation}
SO(4) contains the pair ({\ref{Sdef}) and another one given by
\begin{equation}
A^2=\frac{1}{2}\left(
\sum_{|\mu|\leq\Omega/2}(-)^{j-\mu}a^{\mu}a^{-\mu}-
\sum_{|\mu|>\Omega/2}(-)^{j-\mu}a^{\mu}a^{-\mu}
\right)\ .
\label{Ddef}
\end{equation}
In the notation of Eq.~(\ref{bifer}) we can write
\begin{eqnarray}
\chi^1_{\mu\nu} & = & \left\{
\begin{array}{ll}
(-)^{j-\mu} & {\rm for\ } \nu=-\mu,\\
0           & {\rm for\ } \nu\neq -\mu,
\end{array}\right.
\label{Sdag}
\\
\chi^2_{\mu\nu} & = & \left\{
\begin{array}{ll}
(-)^{j-\mu}   & {\rm for\ } \nu=-\mu, |\mu|\leq\Omega/2,\\
-(-)^{j-\mu} & {\rm for\ } \nu=-\mu, |\mu|>\Omega/2,\\
0             & {\rm for\ } \nu\neq -\mu
\end{array}\right.
\label{Ddag}
\end{eqnarray}
and from Eq.~(\ref{norma}) we get $g=\Omega$.

$A^1$ and $A_1$ together with the commutator $[A_1,A^1]=\Omega-N$
(where $N=a^{\mu}a_{\mu}$ is the real-fermion number operator)
close the SU(2) algebra with the only structure constant
$c^{11}_{11}=2$. In the extended case we introduce an operator
\begin{equation}
Q=\sigma_{\nu}^{\mu}a^{\nu}a_{\mu}\ ,
\label{Q}
\end{equation}
where $\sigma_{\nu}^{\mu}=\pm\delta_{\nu}^{\mu}$ with the
upper (lower) sign valid for $|\mu|\leq\Omega/2$ ($|\mu|>
\Omega/2$). The bifermion operators together with $N$ and $Q$
then close the SO(4) algebra, $[A_1,A^1]=[A_2,A^2]=\Omega-N,
[A_1,A^2]=-Q$, with the following structure constants:
\begin{eqnarray}
 c_{11}^{11}=c_{22}^{22}=c_{22}^{11}=c_{11}^{22}
=c_{12}^{12}=c_{21}^{21}=c_{21}^{12}=c_{12}^{21} & = & 2\ ,
\nonumber\\
c_{11}^{12}=c_{22}^{21}=c_{22}^{12}=c_{11}^{21}
=c_{12}^{11}=c_{21}^{22}=c_{21}^{11}=c_{12}^{22} & = & 0\ .
\end{eqnarray}

It is evident that by dividing the fermionic space into
two subspaces, the first one spanned by single-fermion states
with $|\mu|\leq\Omega/2$ and the other by states with
$|\mu|>\Omega/2$, the SO(4) algebra can be decomposed
into a tensor product of two independent SU(2) algebras.
Accordingly define the following transformation of bifermion
operators:
\begin{eqnarray}
A^< & = & \frac{1}{2}\,(A^1+A^2)=
\frac{1}{2}\sum_{|\mu|\leq\Omega/2}(-)^{j-\mu}a^{\mu}a^{-\mu}\ ,
\label{rota1}\\
A^> & = & \frac{1}{2}\,(A^1-A^2)=
\frac{1}{2}\sum_{|\mu|>\Omega/2}(-)^{j-\mu}a^{\mu}a^{-\mu}\ .
\label{rota2}
\end{eqnarray}
Both $A^<$ and $A^>$ are just the $S^{\dagger}$-type bifermions
in the respective subspaces, cf. Eq.~({\ref{Sdef}}). We have
$[A_<,A^<]=(\Omega/2-N_<)$, $[A_>,A^>]=(\Omega/2-N_>)$,
and $[A_<,A^>]=0$, where $N_<$ and $N_>$ are fermion number
operators associated with both the subspaces: $N_<=(N+Q)/2$,
$N_>=(N-Q)/2$. The only nonzero structure constants are
$c^{<<}_{<<}=c^{>>}_{>>}=2$. Let us note that the new bifermion
states are not generally normalized to a common factor:
$\matr{0}{A_<A^<}{0}=\Omega_<$, $\matr{0}{A_>A^>}{0}=\Omega_>$
with $\Omega_<=\Omega_>=\Omega/2$ for $\Omega$ even, but
$\Omega_<=(\Omega+1)/2$, $\Omega_>=(\Omega-1)/2$ for $\Omega$
odd. A common normalization for odd $\Omega$ would introduce
some additional factors which we skip here for the sake of
simplicity.

\subsection{Mapping of the even sector}
\label{examplesb}

By the straightforward application of Eqs.~(\ref{opmap1})
and (\ref{simi0}) we get
\begin{equation}
A^1\ \longmapsto \ B^1(\Omega-N_{\rm BF})
\label{bisu2}
\end{equation}
for the SU(2) algebra and
\begin{eqnarray}
A^1\ & \longmapsto & \ B^1(\Omega-N_{\rm BF})
-B^2({\cal Q}+B^iB_{i'})
\ ,\label{bi1so4}\\
A^2\ & \longmapsto & \ B^2(\Omega-N_{\rm BF})-B^1
({\cal Q}+B^iB_{i'})
\ ,\label{bi2so4}
\end{eqnarray}
for the SO(4) algebra. Here we introduce boson creation and
annihilation operators $B^i$ and $B_i$ with $i=1$ for SU(2)
and $i=1,2$ for SO(4). We also define $N_{\rm B}=B^iB_i$,
${\cal N}=\alpha^{\mu}\alpha_{\mu}$, $N_{\rm BF}=N_{\rm B}+
{\cal N}$, and ${\cal Q}=\sigma_{\nu}^{\mu}\alpha^{\nu}
\alpha_{\mu}$. In the SO(4) case the summation convention is
used such that $B^iB_{i'}$ stands for $B^1B_2+B^2B_1$. From
Eq.~(\ref{comap}) it follows that
\begin{equation}
N\ \longmapsto\ {\cal N}+2N_{\rm B}
\label{iman}
\end{equation}
for both the SU(2) and SO(4), and
\begin{equation}
Q\ \longmapsto\ {\cal Q}-2B^iB_{i'}
\label{iq}
\end{equation}
for the SO(4).

Instead of $A^1$ and $A^2$ we can also map the bifermions from
Eqs.~(\ref{rota1}) and (\ref{rota2}). The result is then
\begin{eqnarray}
A^<\ & \longmapsto & \ B^<(\Omega/2-N_{{\rm BF}<})
\ ,\label{bi1so4'}\\
A^>\ & \longmapsto & \ B^>(\Omega/2-N_{{\rm BF}>})
\ ,\label{bi2so4'}\\
N_<\ & \longmapsto & \ {\cal N}_<+2N_{{\rm B}<}
\ ,\label{iman'}\\
N_>\ & \longmapsto & \ {\cal N}_>+2N_{{\rm B}>}
\ ,\label{iman''}
\end{eqnarray}
where $N_{{\rm B}<}=B^<B_<$, $N_{{\rm B}>}=B^>B_>$, ${\cal N}_<=
({\cal N}+{\cal Q})/2$, ${\cal N}_>=({\cal N}-{\cal Q})/2$,
$N_{{\rm BF}<}=N_{{\rm B}<}+{\cal N}_<$, and $N_{{\rm BF}>}=
N_{{\rm B}>}+{\cal N}_>$. Both SO(4) results,
Eqs.~(\ref{bi1so4})--(\ref{iq}) and (\ref{bi1so4'})--(\ref{iman''}),
can be combined using the bosonic counterpart of the transformation
in Eqs.~(\ref{rota1}) and (\ref{rota2}),
\begin{eqnarray}
\begin{array}{ll}
B^<=B^1+B^2\ ,              & B^>=B^1-B^2\ ,\\
B_<=\frac{1}{2}(B_1+B_2)\ , & B_>=\frac{1}{2}(B_1-B_2)\ ,
\end{array}
\label{rota'}
\end{eqnarray}
which results from the linearity of mapping. It should be noted
that the new boson creation and annihilation operators in
Eq.~(\ref{rota'}), unlike $B^i$ and $B_i$ with $i=1,2$, are not
related by the hermitian conjugation---a result of non-unitarity
of the mapping. If, in contrary, $B^{\bullet}$ and $B_{\bullet}$
with $\bullet=<$ and $>$ were chosen to be hermitian conjugeted,
the same would not hold for $B^i$ and $B_i$.

The mapping of the most general one- plus two-body hamiltonian
(\ref{Hamr}), evaluated for the two algebras under discussion, yields
\begin{equation}
H\ \longmapsto\ (u+\Omega w^1_1)-w^1_1{\cal N}+
(v^1_1-2w^1_1)N_{\rm B}+v^1_1N_{\rm B}(\Omega-
N_{\rm BF})
\label{Hamsu2}
\end{equation}
for SU(2) and
\begin{eqnarray}
H\ \longmapsto\ & & (u+\Omega w^i_i)-w^i_i{\cal N}-
w^i_{i'}{\cal Q}+(v^i_i-2w^i_i)N_{\rm B}+(v^j_{j'}+
2w^j_{j'})B^iB_{i'}
\nonumber\\
& & +v^j_iB^iB_j(\Omega-N_{\rm BF})-
(v^1_2B^1B_1+v^2_1B^2B_2+v^1_1B^2B_1+v^2_2B^1B_2)
({\cal Q}+B^iB_{i'})
\label{Hamso4}
\end{eqnarray}
for the SO(4). We note that whereas the mapped SU(2)
hamiltonian is manifestly hermitian, the SO(4) hamiltonian is not,
because of its last term.

\subsection{Similarity transformations and mapping of
the odd sector}
\label{examplesc}

Let us finally focus on the form of single-fermion images for
both algebras. The similarity transformation (\ref{simi2})
depends on the form of the Casimir operator ${\cal C}_{\rm F}$.
For the SU(2) algebra we can introduce the seniority quantum
number $v$ \cite{Greiner} such that
\begin{equation}
{\cal C}_{\rm F}\equiv{\cal A}^1{\cal A}_1=\frac{1}{4}
\left[(\Omega-v)(\Omega+2-v)-(\Omega-{\cal N})
(\Omega+2-{\cal N})\right]
\label{senior}
\end{equation}
in the seniority eigenbasis. Because ${\cal A}^1B_1$ does not
change $v$ (the number of fermions {\em not\/} coupled in pairs),
the first term in Eq.~(\ref{senior}) does not contribute in
Eq.~(\ref{simi2}) and one gets \cite{Navratil2}
\begin{equation}
{\cal C}_{\rm F}-{\hat{\cal C}}_{\rm F}\ \longrightarrow\
\frac{1}{2}\left({\cal N}-{\hat{\cal N}}\right)
\left(\Omega+1-\frac{{\cal N}+{\hat{\cal N}}}{2}\right)\ .
\end{equation}
The operators $S_k$ in Eq.~(\ref{s1k}) thus read as
\begin{equation}
S_k=\frac{1}{k!}({\cal A}^1B_1)^k
\frac{(\Omega-{\cal N}-k)!}{(\Omega-{\cal N})!}\ ,
\label{sksu2}
\end{equation}
which is equivalent to the known expression \cite{Navratil2}
\begin{equation}
S_{\rm B}^{-1}=\frac
{\bigl(\Omega-\frac{{\cal N}+{\hat{\cal N}}}{2}\bigr)!}
{(\Omega-{\hat{\cal N}})!}\exp({\cal A}^1B_1)_{\wedge}
\label{simisu2}
\end{equation}
[cf. Eq.~(\ref{sumpos2}) in Appendix A].
It is instructive to note that the above expressions for the
similarity transformation converge under limited conditions
only. Consider the case of ${\cal N}\leq\Omega$. Then we see
that the expression in Eq.~(\ref{sksu2}) diverges for
$\Omega-{\cal N}+1\leq k\leq N_{\rm B}$ (the upper limit follows
from the fact that $S_k$ gives just zero if it attempts to annihilate
too many bosons). So the divergence problems are avoided if
\begin{equation}
{\cal N}+N_{\rm B}=\frac{N+{\cal N}}{2}\leq \Omega\ .
\label{conver}
\end{equation}
Beyond the validity of Eq.~(\ref{conver}) the forms
(\ref{sksu2}) and (\ref{simisu2})
of the similarity transformation is invalid and another derivation would
be
required [see the remark below Eq.~(\ref{pert}) in Appendix A].

In the SO(4) case the construction of a similarity transformation
turns out to be more difficult as the
denominator in Eq.~(\ref{simi2}) cannot be expressed as a function
of ${\cal N}$. We can, however, use the SU(2)$\otimes$SU(2) type
of mapping, Eqs.~(\ref{bi1so4'})--(\ref{rota'}), for which the
analogy with the single SU(2) case can be fully exploited.
The Casimir operator ${\cal C}_{\rm F}=2({\cal A}^<{\cal A}_<+
{\cal A}^>{\cal A}_>)$ then reads as a sum of two terms of the
form (\ref{senior}). Again, seniorities corresponding to both
subspaces are not affected by ${\cal A}^iB_i=2({\cal A}^<B_<+
{\cal A}^>B_>)$ and the following substitution can be used within
Eq.~(\ref{simi2}):
\begin{eqnarray}
{\cal C}_{\rm F}-{\hat{\cal C}}_{\rm F}\ \longrightarrow\
& & \left({\cal N}_<-{\hat{\cal N}}_<\right)
\left(\Omega_<+1-\frac{{\cal N}_<+{\hat{\cal N}}_<}{2}\right)
\nonumber\\
& + & \left({\cal N}_>-{\hat{\cal N}}_>\right)
\left(\Omega_>+1-\frac{{\cal N}_>+{\hat{\cal N}}_>}{2}\right)
\ .\label{denoso4}
\end{eqnarray}
From this expression we find that
\begin{equation}
S_k=\sum_{k_<+k_>=k}\frac{1}{k_<!k_>!}
({\cal A}^<B_<)^{k_<}({\cal A}^>B_>)^{k_>}
\frac{(\Omega_<-{\cal N}_<-k_<)!(\Omega_>-{\cal N}_>-k_>)!}
{(\Omega_<-{\cal N}_<)!(\Omega_>-{\cal N}_>)!}\ ,
\label{skso4}
\end{equation}
with the summation going from $k_<,k_>=0$ to $k$. In analogy
with Eq.~(\ref{simisu2}) we also have
\begin{equation}
S_{\rm B}^{-1}=\frac{\bigl(\Omega_<-\frac{{\cal N}_<+
{\hat{\cal N}}_<}{2}\bigr)!}{(\Omega_<-{\hat{\cal N}}_<)!}
\frac{\bigl(\Omega_>-\frac{{\cal N}_>+{\hat{\cal N}}_>}{2}
\bigr)!}{(\Omega_>-{\hat{\cal N}}_>)!}\exp({\cal A}^<B_<+
{\cal A}^>B_>)_{\wedge}\ .
\label{simiso4}
\end{equation}
For ${\cal N}_<\leq\Omega_<$ and ${\cal N}_>\leq\Omega_>$
the convergence of Eq.~(\ref{skso4}) requires, in analogy
with Eq.~(\ref{conver}), $N_{{\rm B}<}\leq\Omega_<-{\cal N}_<$
and $N_{{\rm B}>}\leq\Omega_>-{\cal N}_>$.

Now we can evaluate the similarity transformation of the
images of single-fermion annihilation and creation operators.
We already know that in general the resulting series
contains terms changing the number of ideal fermions by
$\Delta{\cal N}=-1,+1,+3,\dots$.  In the SU(2) case, however,
all terms with $\Delta{\cal N}\geq +3$ vanish.
Indeed, it can be shown that with $S_k$ from Eq.~(\ref{sksu2})
we have
\begin{equation}
[\alpha_{\mu},S_k]-S_{k-1}[\alpha_{\mu},S_1]=0
\label{vanish1}
\end{equation}
for $k=2,3,\dots$, which together with Eq.~(\ref{recur})
implies that $X_k=0$ for $k\geq 2$ with $X=\alpha_{\mu}$
being the bare image of the annihilation operator.
In the bare image of the creation operator, $X=
\alpha^{\mu}+B^1{\tilde{\alpha}_{\mu}}\equiv X'+X''$,
where ${\tilde\alpha}_{\mu}=(-)^{j-\mu}\alpha_{-\mu}$,
the first term changes the ideal-fermion number by $+1$
and the second by $-1$. The condition for the cancellation
of $\Delta{\cal N}\geq +3$ terms in the transformed image
therefore reads as $X'_k=-X''_{k+1}$, i.e.,
\begin{equation}
[\alpha^{\mu},S_k]+[B^1{\tilde\alpha}_{\mu},S_{k+1}]-
S_k[B^1{\tilde\alpha}_{\mu},S_1]=0
\label{vanish2}
\end{equation}
for $k=1,2,3,\dots$. Again, it can be proven from
Eq.~(\ref{sksu2}) that Eq.~(\ref{vanish2}) is valid. For
the transformed SU(2) single-fermion
images we finally obtain \cite{Navratil2}:
\begin{eqnarray}
a_{\mu}\ & \longmapsto & \
\alpha_{\mu}+{\tilde\alpha}^{\mu}B_1\frac{1}{\Omega-{\cal N}}
+{\cal A}^1B_1\alpha_{\mu}\frac{1}{(\Omega+1-{\cal N})
(\Omega-{\cal N})}\ ,
\label{single1su2}\\
a^{\mu}\ & \longmapsto & \
B^1{\tilde\alpha}_{\mu}+
\alpha^{\mu}\,\frac{\Omega-N_{\rm BF}}{\Omega-{\cal N}}
-{\cal A}^1{\tilde\alpha}_{\mu}
\frac{\Omega-N_{\rm BF}}{(\Omega+1-{\cal N})
(\Omega-{\cal N})}.
\label{single2su2}
\end{eqnarray}
[In analogy to ${\tilde\alpha}_{\mu}$ we define
${\tilde\alpha}^{\mu}=(-)^{j-\mu}\alpha^{-\mu}$.]

To derive the single-fermion images in the SO(4) case, one
first shows that Eqs.~(\ref{vanish1}) and (\ref{vanish2})
are again fulfilled with $S_k$ from Eq.~(\ref{skso4}), if
$B^1$ in Eq.~(\ref{vanish2}) is replaced by $B^{\bullet}
\equiv B^<$ or $B^>$ according to whether $|\mu|\leq\Omega/2$
or $|\mu|>\Omega/2$, respectively. This means that the series
for the transformed single-fermion images
$S_{\rm B}\alpha_{\mu}S_{\rm B}^{-1}$ and
$S_{\rm B}(\alpha^{\mu}+B^{\bullet}{\tilde\alpha}_{\mu})
S_{\rm B}^{-1}$ both terminate at the terms with
$\Delta{\cal N}=+1$. We thus obtain
\begin{eqnarray}
a_{\mu}\ & \longmapsto & \
\alpha_{\mu}+{\tilde\alpha}^{\mu}B_{\bullet}
\frac{1}{\Omega_{\bullet}-{\cal N}_{\bullet}}
+{\cal A}^{\bullet}B_{\bullet}\alpha_{\mu}
\frac{1}{(\Omega_{\bullet}+1-{\cal N}_{\bullet})
(\Omega_{\bullet}-{\cal N}_{\bullet})}\ ,
\label{single1so4}\\
a^{\mu}\ & \longmapsto & \
B^{\bullet}{\tilde\alpha}_{\mu}
+\alpha^{\mu}\,\frac{\Omega_{\bullet}-N_{{\rm BF}\bullet}}
{\Omega_{\bullet}-{\cal N}_{\bullet}}
-{\cal A}^{\bullet}{\tilde\alpha}_{\mu}
\frac{\Omega_{\bullet}-N_{{\rm BF}\bullet}}
{(\Omega_{\bullet}+1-{\cal N}_{\bullet})
(\Omega_{\bullet}-{\cal N}_{\bullet})}\ ,
\label{single2so4}
\end{eqnarray}
where the bullet $\bullet$ stands for $>$ or $<$, depending on
which subspace $\mu$ belongs to. Eqs.~(\ref{single1so4}) and
(\ref{single2so4}) are direct analogues of the single-fermion
images in the SU(2) case, cf. Eqs.~(\ref{single1su2}) and
(\ref{single2su2}).

\subsection{Fermion and bifermion transfer matrix elements}
\label{examplesd}

To demonstrate the utility of the results derived above, we calculate
matrix elements of single-fermion and fermion-pair transfer operators
using the ideal boson-fermion images. We start with the SU(2) case,
where we consider the following three normalized fermionic states:
\begin{eqnarray}
\ket{\psi_0} & = & C_0(A^1)^{N/2}\ket{0}\ ,
\nonumber\\
\ket{\psi_1} & = & C_1a^{\mu}(A^1)^{N/2}\ket{0}\ ,
\label{psisu2}\\
\ket{\psi_2} & = & C_2(A^1)^{N/2+1}\ket{0}
\nonumber
\end{eqnarray}
($N$ or $N+2$ are even numbers of paired fermions). The matrix
elements of the single-fermion and fermion-pair transfer operators
between these states depend just on the normalization constants
$C_0$, $C_1$, and $C_2$ and one readily finds
\begin{eqnarray}
\matr{\psi_1}{a^{\mu}}{\psi_0}=\matr{\psi_0}{a_{\mu}}{\psi_1}
& = & \frac{C_0}{C_1}=\sqrt{\frac{2\Omega-N}{2\Omega}}\ ,
\label{mat1}\\
\matr{\psi_2}{A^1}{\psi_0}=\matr{\psi_0}{A_1}{\psi_2}
& = & \frac{C_0}{C_2}=\frac{1}{2}\sqrt{(2\Omega-N)(N+2)}\ .
\label{mat2}
\end{eqnarray}

The results given in Eqs.~(\ref{mat1}) and (\ref{mat2}) are
reproduced in the ideal space, using Eq.~(\ref{matrixele})
with the single-fermion and fermion-pair images
$\overline{a_{\mu}}$ from Eq.~(\ref{single1su2}),
$\overline{a^{\mu}}$ from Eq.~(\ref{single2su2}),
$\overline{A_1}=B_1$, and $\overline{A^1}$ from Eq.~(\ref{bisu2}).
The left and right ideal states corresponding to
Eq.~(\ref{psisu2}) read as follows,
\begin{eqnarray}
\keti{\psi^{\rm R}_0} & = & C_0(\overline{A^1})^{N/2}
\keti{0}=C^{\rm R}_0(B^1)^{N/2}\keti{0}\ ,
\nonumber\\
\brai{\psi^{\rm L}_0} & = & \brai{0}\,(\overline{A_1})^{N/2}
C^*_0=\brai{0}\,(B_1)^{N/2}C^{\rm L}_0\ ,
\nonumber\\
\keti{\psi^{\rm R}_1} & = & C_1\overline{a^{\mu}}
(\overline{A^1})^{N/2}\keti{0}=C^{\rm R}_1\alpha^{\mu}
(B^1)^{N/2}\keti{0}\ ,
\label{LRpsisu2}\\
\brai{\psi^{\rm L}_1} & = & \brai{0}\,(\overline{A_1})^{N/2}
\overline{a_{\mu}}C^*_1=\brai{0}\,(B_1)^{N/2}\alpha_{\mu}
C^{\rm L}_1\ \dots
\nonumber
\end{eqnarray}
(the images of $\ket{\psi_2}$ are analogous to the ones of
$\ket{\psi_0}$). It is clear that the coefficients $C^{\rm R}_i$
and $C^{\rm L}_i$ carry information on the specific construction
of the images in Eq.~(\ref{LRpsisu2}) from the real states,
in particular information on the fermionic normalization
constants. This seems to undermine the practical implementation
of the mapping
procedure because once the mapping of a particular algebra has
been established,
one certainly wants to be able to perform all the
calculations solely on the ideal boson-fermion level. Here we
come to the reason why Eq.~(\ref{matrixele}) is more convenient
from the phenomenological viewpoint than the seemingly simpler
identity $\matr{\psi_1}{O}{\psi_2}=\matri{\psi_1^{\rm L}}
{\overline O}{\psi_2^{\rm R}}$: with Eq.~(\ref{matrixele})
the results depend just on the products
$C^{\rm L}_iC^{\rm R}_i$ (with $i=1,2$) that can be easily
determined using only the {\em bosonic\/} (ideal) normalization
condition $\scali{\psi^{\rm L}_i}{\psi^{\rm R}_i}=1$. We see
therefore that the matrix elements (\ref{mat1}) and
(\ref{mat2}) can be calculated on the purely
{\em phenomenological\/} level---i.e., using only ideal
boson-fermion states with no explicit reference to their
real fermionic ancestors---provided that we know
microscopically-based ideal images of (bi)fermion creation
and annihilation operators. This again emphasizes the
importance of the construction carried out in Secs.~\ref{single}
and \ref{examplesc}.

In the SO(4) case one can proceed in a close analogy with SU(2).
It turns out that it is much easier to work in the collective
basis created by pairs $A^<$ and $A^>$ rather than $A^1$ and
$A^2$. We thus define fermionic states
\begin{eqnarray}
\ket{\psi_0} & = & C_0(A^>)^{N/2-k}(A^<)^k\ket{0}\ ,
\nonumber\\
\ket{\psi_1} & = & C_1a^{\mu}(A^>)^{N/2-k}(A^<)^k\ket{0}\ ,
\nonumber\\
\ket{\psi_2} & = & C_2(A^>)^{N/2-k}(A^<)^{k+1}\ket{0}\ ,
\label{psiso4}\\
\ket{\psi'_2} & = & C'_2(A^>)^{N/2-k+1}(A^<)^k\ket{0}\ .
\nonumber
\end{eqnarray}
Note that we now have two possibilities, $\ket{\psi_2}$ and
$\ket{\psi'_2}$, of building a paired ($N$+2)-fermion states
from $\ket{\psi_0}$. With the aid of the left and right ideal
states corresponding to Eq.~(\ref{psiso4}) [similar to those
in Eq.~(\ref{LRpsisu2})] and the operator images in
Eqs.~(\ref{bi1so4'}), (\ref{bi2so4'}), (\ref{single1so4}),
and (\ref{single2so4}), it is now simple to verify that
Eq.~(\ref{matrixele}) yields
\begin{eqnarray}
\matr{\psi_1}{a^{\mu}}{\psi_0}=\matr{\psi_0}{a_{\mu}}{\psi_1}=
\left\{
\begin{array}{ll}
\sqrt{\frac{\Omega_<-k}{\Omega_<}} &
{\ \rm for\ }|\mu|\leq\Omega/2\ , \\
\sqrt{\frac{2\Omega_>-N+2k}{2\Omega_>}} &
{\ \rm for\ }|\mu|>\Omega/2
\end{array}
\right.
\label{matt1}
\end{eqnarray}
and
\begin{eqnarray}
\matr{\psi_2}{A^<}{\psi_0}=\matr{\psi_0}{A_<}{\psi_2}
& = & \sqrt{\frac{(\Omega-2k)(k+1)}{2}}\ ,
\label{matt2}\\
\matr{\psi'_2}{A^>}{\psi_0}=\matr{\psi_0}{A_>}{\psi'_2}
& = & \frac{1}{2}\sqrt{(\Omega-N+2k)(N-2k+2)}\ ,
\label{matt3}\\
\matr{\psi'_2}{A^<}{\psi_0}=\matr{\psi_0}{A_<}{\psi'_2} & = &
\matr{\psi_2}{A^>}{\psi_0}=\matr{\psi_0}{A_>}{\psi_2}=0\ .
\label{matt4}
\end{eqnarray}
These results can be checked by evaluating the fermionic
normalization constants. Let us point out that the
calculation would be much more involved if we chose
to use the collective basis created by $A^1$ and $A^2$. Since
ideal images of these operators contain both $B^1$ and $B^2$
[see Eqs.~(\ref{bi1so4}) and (\ref{bi2so4})] the mapped collective
states (right images) would combine various numbers of type-1
and 2 bosons (only the total boson number being constant).

\section{Conclusions}

We investigated various aspects of the generalized Dyson mapping
that transforms fermionic shell-model superalgebras into the ideal
boson-fermion space \cite{Dobaczewski1,Navratil1,Navratil2}.
The main motivation for this review was the recent experimental
verification \cite{Metz,Metz2,Groger} of the phenomenological
boson-fermion supersymmetric model \cite{Isacker} and the resulting
renewed interest in its microscopic foundations. Along with
presenting some particular new results we found it useful also
to summarize in a compact form the main principles of the underlying
mathematical formalism and the hurdles that remain.

While in the standard Dyson mapping only the collective algebra
of fermion pair operators is transformed into the ideal space,
yielding a set of purely bosonic images, the generalized Dyson
mapping transforms also the single-fermion creation and
annihilation operators, i.e., the whole superalgebra defined
in Sec.~\ref{superalgebra}. As a result, ideal-fermion operators
enter the images of physical observables in addition to the boson
operators. The mapping procedure outlined here makes use of the
generalized Usui operator (\ref{Urea}), which has the advantage
of providing in a relatively straightforward manner a first set of
simple formulas---Eqs.~(\ref{opmap1})--(\ref{comap})---for the
images of the operators involved in the superalgebra. However,
it turns out that some additional transformations are needed to
accomplish the physically motivated
bosonisation and unitarity of the mapping. The general
form of these transformations was discussed in
Sec.~{\ref{simila}, while in Appendix~A we provided technical
insight into the formalism used for their derivation.

We studied in particular the \lq\lq bosonisation\rq\rq\
similarity transformations, see Eqs.~(\ref{simi2})--(\ref{s2k}).
Without these transformations, the main aim of the
Mapping---replacement of the fermionic correlations by simpler
bosonic correlations---would not be achieved, since all
fermion-fermion interactions would be exactly reproduced
in the ideal-particle space. The action of the bosonisation
similarity transformation on a general operator was determined
in the expanded form of Eqs.~(\ref{xxx})--(\ref{recur}). These
expressions represent a new result compared to previous work on
this subject. However, to use them in general for deriving closed
expressions might still be elusive unless the Casimir
operator ${\cal C}_{\rm F}$
of the ideal-fermion core algebra turns out to depend solely on
the number $\cal N$ of ideal fermions in the whole space or
its specific subspaces. If this condition is fulfilled, the
calculations can be carried out further and one derives, e.g.,
the explicit form of transformed single-fermion images in
Eqs.~(\ref{ani}) and (\ref{cre}). These results are already of
importance
to hint at suitable expressions for nucleon transfer operators
in phenomenological supersymmetric models.

A particularly interesting question, related directly to the
microscopic justification of phenomenological supersymmetric
models, concerns the conservation of the total number of ideal
particles (fermions plus bosons). It was shown in
Sec.~\ref{conserva} that this number is indeed a natural
integral of motion if the even sector of the mapped superalgebra
is chosen properly, i.e., so that it fully represents the real
fermionic hamiltonian. From the point
of view of the generalized Dyson mapping, the origin of the
phenomenological U($M/2\Omega$) dynamical superalgebra seems
to have a sound microscopic basis. Since numbers of ideal
bosons and fermions
turn out to be conserved separately, the present method also
advocates the decomposition of the phenomenological dynamical
superalgebra into the product of bosonic and fermionic algebras
in the first step of the relevant dynamical-symmetry chains
\cite{Isacker}. However, the realization of truly supersymmetric
predictions that are not specifically connected with dynamical
symmetries, as discussed in Refs.\cite{Jolos,Jolos2,Jolos3},
is not excluded.

To illustrate the general technique outlined in this paper, we
investigated in Sec.~\ref{examples} concrete examples of mapping
the SU(2) and SO(4) collective superalgebras. The results for the
seniority SU(2) model were derived earlier \cite{Navratil2}, but
we reconsidered them from a more general point of view and to
facilitate the analysis of the SO(4) case, originally discussed
by Kaup and Ring \cite{Kaup}. As the SO(4) algebra can also be
written as the product SU(2)$_<\otimes$SU(2)$_>$, it provides an
interesting insight into the link between boson images of the two
different realizations, as, e.g., in Eq.~(\ref{rota'}). Both the
SU(2) and SO(4) models exemplify the relative simplicity of the
Dyson mapping which follows from \lq\lq bare\rq\rq\ operator
images, while they also point to technical difficulties
associated with similarity transformations. For the specific
superalgebras studied here the bosonisation similarity transformation
leads to the closed expressions for single-fermion images given
in Eqs.~(\ref{single1su2})--(\ref{single2so4}). In more general
cases, however, the transformed images may involve more complicated
series, where convergence becomes an issue. This problem must
be ovecome for an optimal comparison with the phenomenological
framework. Note also that unlike the models considered here, the
dynamical definition of collective fermion pairs (bosons)
requires attention beyond the algebraic definitions (see, e.g.,
Ref.\cite{Klein}).

One of the main remaining obstacles in the quantitative microscopic
analysis of phenomenological supersymmetric models is associated with
the nonunitarity of the generalized Dyson mapping. While this property
obscures some aspects of a direct comparison with phenomenology, we
also documented that on the matrix-element level the formalism can
already be implemented in a way which closely resembles the
phenomenological application. This was illustrated through the use
of Eq.~(\ref{matrixele}) leading to the examples in
Sec.~\ref{examplesd}.

\acknowledgements{This work was supported by the S.A. National
Research Foundation under grants GUN 2047181 and GUN 2044653 and
partly by the Grant Agency of Czech Republic under grant
202/99/1718. P.C. would also like to acknowledge the hospitality
and enjoyable working conditions provided by the ITP at the
University of Stellenbosch.}

\appendix

\section{Positional Operator Calculus}

Consider an operator $O=O'+P$ where $O$ and $O'$ are isospectral
and denote $O\ket{\psi_i}=o_i\ket{\psi_i}$ and $O'\ket{\psi'_i}=
o_i\ket{\psi'_i}$. In general, $O$ does not have to be hermitian,
so that the $\ket{\psi_i}$ are not necessarily orthogonal, while
$O'$ {\em is\/} hermitian, implying that its eigenvectors
$\ket{\psi'_i}$ form an orthonormal basis. The transformation
connecting the two sets of eigenvectors, $S\ket{\psi_i}=
\ket{\psi'_i}$, transforms away the $P$-term of $O$, i.e.,
$S(O'+P)S^{-1}=O'$. This is the property of similarity
transformations required in Sec.~\ref{simila}. The form of
$S^{-1}$ can be determined from the ordinary perturbative
series expressing $\ket{\psi_i}$ in terms of $\ket{\psi'_i}$
(with $P$ treated as a perturbation). The isospectrality
condition is often guaranteed by the fact that $P$ has the
upper (lower) off-diagonal block structure in the basis
$\ket{\psi'_i}$. In that case $\matr{\psi'_i}{P}{\psi'_i}=0$
and the expansion reads as follows:
\begin{equation}
\ket{\psi_i}=\sum_{k=0}^{\infty}\left(\frac{1}{o_i-O'}P\right)^k
\ket{\psi'_i}
\label{pert}
\end{equation}
Note that the terms with $k>0$ are to be evaluated only in case
of $[O',P]\neq 0$, otherwise they are equal to zero. It must be
stressed that Eq.~(\ref{pert}) is derived using the perturbation
theory for nondegenerated cases. Its applicability is thus not
quite universal and the convergence conditions should be determined
in each particular case.

In fact, Eq.~(\ref{pert}) defines the action of $S^{-1}$ on any
vector via its expansion in the eigenbasis $\ket{\psi'_i}$. To avoid
the explicit reference to the basis, the idea of positional operators
was introduced in Refs.~\cite{Kim,Geyer1}. Eq.~(\ref{pert}) can be
rewritten as
\begin{equation}
S^{-1}=\sum_{k=0}^{\infty}\sum_i\left(\frac{1}{o_i-O'}P\right)^k
\ket{\psi'_i}\bra{\psi'_i}=\sum_{k=0}^{\infty}\left(\frac{1}
{{\hat O'}-O'}P\right)^k_{\wedge}\ ,
\label{posiA}
\end{equation}
where the hat above the operator $O'$ means that in the expansion of
each term on the r.h.s. of Eq.~(\ref{posiA}) this operator must be
evaluated at the position indicated by \lq\lq $_{\wedge}$\rq\rq.

One can develop a general calculus suitable for handling expressions
like the one in Eq.~(\ref{posiA}). In fact, any hatted operator is
treated as an ordinary $c$-number during the evaluation, i.e., it
may freely travel to any place as far as its true position is
marked. Any part of $O'$ which commutes with $P$ cancels with the
corresponding part of ${\hat O'}$, so that $O'$ in Eq.~(\ref{posiA})
can be replaced by any operator $C$ that satisfies $[C,P]=[O',P]$.
We thus arrive at Eq.~(\ref{posi}). With no further assumption
upon commutation relations between the operators involved, the
evaluation of terms like $f({\hat C}-C)AB_{\wedge}$ must
unavoidably deal with the decomposition of the function $f(x)$
into a series, which usually leads to rather complicated expressions.
For example, for $f(x)=\sum_{n=0}^{\infty}f_nx^n$ one can derive
\begin{equation}
f(C-{\hat C})AB_{\wedge}=f(C-{\hat C})A_{\wedge}B+
\sum_{n=0}^{\infty}f_n\sum_{k=0}^n(-)^{n-k}
\left(\begin{array}{c}n\\k\end{array}\right)
\left([C^k,A][B,C^{n-k}]+A[C^k,B]C^{n-k}\right)\ .
\label{devil}
\end{equation}
However, great simplification can be achieved if $A,B,C$  conform to
commutation relations like $f(c-C)A=Ag(c-C)$, where $g(x)$
and $f(x)$ are some interrelated functions and $c$ an arbitrary
constant.

Let us consider an important special case with $C=aN^2+bN+c$, where
the operator $N$ fulfills the condition $NP=P(N+m)$ (with $m$
a positive integer) and $a,b,c$ are constants. $P$ is a ladder
operator for $N$ and we assume eigenvalues of $N$ within the range
from 0 to $n_{\rm max}>0$. Since $N$ here represents the
fermion-number operator and $n_{\rm max}$ the shell
capacity, the above condition is satisfied if $P$ creates $m$
fermions and $C$ is a quadratic function of $N$. The sum in
Eq.~(\ref{posiA}) terminates at $k_{\rm max}=\lfloor n_{\rm max}/
m\rfloor$. Moreover, for some values of the constants the series can
be formally summed, yielding
\begin{equation}
S^{-1}=\frac{(2{\hat N}+\frac{b}{a})!^m}
{(N+{\hat N}+\frac{b}{a})!^m}\,
\exp{\left(-\frac{a}{m}P\right)}_{\wedge}
\label{sumpos1}
\end{equation}
for $b/a>0$ and
\begin{equation}
S^{-1}=\frac{(-\frac{b}{a}-m-N-{\hat N})!^m}
{(-\frac{b}{a}-2{\hat N})!^m}
\,\exp{\left(\frac{a}{m}P\right)}_{\wedge}
\label{sumpos2}
\end{equation}
for $b/a<-3k_{\rm max}m$. Here, $!^m$ stands for the
\lq\lq factorial over $m$\rq\rq, i.e., $x!^m=x(x-m)(x-2m)\dots
(x\,{\rm mod}\,m)$ for $x>0$ and $x!^m:=1$ for $x\leq 0$. One may
verify Eqs.~(\ref{sumpos1}) and (\ref{sumpos2}) from the relation
${\hat C}-C=({\hat N}-N)[a(N+{\hat N})+b]$, commuting the first
term to the right and the second term to the left. (The constraints
on $b/a$ ensure that the factorial-like terms contain
only positive numbers; otherwise the above formulas can be used in
a restricted subspace only.) Various specific realizations of
Eqs.~(\ref{sumpos1}) and (\ref{sumpos2}) can be found in
Refs.\cite{Navratil1,Navratil2,Kim,Geyer1}. If $C$ cannot be
expressed as a quadratic function of $N$, but $[C,N]=0$ still holds,
one obtains
\begin{equation}
S^{-1}=\exp{\left[-\frac{{\hat N}-N}{m({\hat C}-C)}
P\right]}_{\wedge}\ .
\label{sumpos3}
\end{equation}

\end{document}